\documentclass[]{mn2e}
\usepackage{psfig, epsf, epsfig}
\title[Probing clumpy galaxies]{Probing clumpy pasts of galaxies 
       from AGB stars}
\author[K. Bekki and M.-R. L. Cioni]
       {Kenji Bekki${}^1$\thanks{E-mail: bekki@bat.phys.unsw.edu.au} 
        and M.-R. L. Cioni ${}^2$ \\
       ${}^1$School of Physics, University of New South Wales,
              Sydney 2052, NSW, Australia\\
       ${}^2$ SUPA, School of Physics, University of Edinburgh, 
             IfA, Blackford Hill, Edinburgh EH9 3HJ, UK \\}
  
\begin{document}

\date{Accepted, Received 2005 May 13; in original form }

\pagerange{\pageref{firstpage}--\pageref{lastpage}} \pubyear{2005}

\maketitle

\label{firstpage}

\begin{abstract}

Recent morphological studies of galaxies by the {\it Hubble
Space Telescope (HST)} have revealed that actively star-forming
galaxies at intermediate and high redshifts ($z=0.5-2.0$)
have very clumpy and irregular distributions of stars. 
It is however unclear whether and how these clumpy galaxies evolve 
into the present spiral and elliptical  galaxies with regular shapes.
We here propose that spatial distributions of AGB stars,
probing the  
different mean age and metallicity  of the underlying stellar population,
can provide  vital clues
to the evolution of these clumpy galaxies, in particular, those
at intermediate redshifts.
In order to demonstrate this proposal to be quite promising,
we show the results of test-particle simulations on
the long-term dynamical evolution of unbound groups of AGB stars
(``stellar clumps''),
which correspond to the successors of star-forming clumps
at intermediate redshifts, 
in isolated and interacting galaxies.
We particularly show that azimuthal 
distributions of AGB stars dispersed from stellar clumps 
as a result of gravitational interaction with their host galaxies
can be still inhomogeneous several Gyrs after stellar clump
formation for some models.
We also show that the inhomogeneities
in the azimuthal distributions of dispersed AGB stars
can more quickly disappear in stellar clumps with
larger sizes and higher velocity dispersions. 
These results suggest that if apparently clumpy structures
of galaxies at intermediate redshifts
are due to stars in unbound or weakly bound clusters,
spatial distributions of AGB stars
can have fossil records on  past clumpy structures of galaxies.
As an example,
we  compare the latest observations on
the spatial distributions of AGB stars in 
the Large and Small Magellanic Cloud (LMC and SMC, respectively) 
with the corresponding simulations and thereby
discuss whether the LMC and the SMC had  massive star-forming clumps
in their  outer disks a few to several Gyrs ago.

\end{abstract}


\section{Introduction}

Recent morphological studies of galaxies
at intermediate and high redshifts ($z=0.5-2.0$) have discovered several
kinds of clumpy galaxies, 
such as ``tadpole  galaxies'',  ``clump-clusters'',
and ``chain galaxies'' (e.g., Cowie et al. 1995; 
van den Bergh et al. 2006;
Reshetnikov et al. 2003;
Conselice et al. 2005;
Elmegreen et al. 2004a, b, 2005).
Recent numerical and theoretical works have suggested that
dynamical evolution
of massive star-forming  clumps in these  clumpy galaxies
are important for the formation of bulges,
thick disks, and their exponential profiles 
in the present galaxies  (e.g., Noguchi 1998).
However observational studies on structural and kinematical
properties of {\it the present galaxies} have not yet
revealed evolutionary links between these distant clumpy galaxies
and the present normal Hubble type ones.

Asymptotic Gian Branch (AGB) stars have been considered
to be useful indicators of galaxy structures and kinematics,
because they are easily noticed owing to their bright
magnitudes and widely  distributed  across galaxies
(e.g., van der Marel et al. 2002; Cioni \& Habing 2003).
Recent studies on structural properties 
of nearby galaxies 
based on spatial distributions of AGB stars
(e.g., LMC, SMC, M33, and NGC 6822)
have revealed clumpy and asymmetric  distributions of
stars for age and metallicity ranges
(e.g., Cioni \& Habing 2003; Cioni et al. 2006a, b).
For example, Cioni et al. (2006b) found that (1)
the azimuthal distributions of stars with mean ages of 2.0-10.6 Gyr
in the SMC
are quite inhomogeneous and (2) the distributions
depend strongly on ages and metallicities of stars.
Although physical origins of the observed inhomogeneities
in azimuthal distributions of stars 
remain unclear,  
it might well be  reasonable to consider that
AGB stars, which evolve from   $0.8-8 {\rm M}_{\odot}$ 
stars and therefore are  stellar populations suitable for investigating
long-term structural  evolution of galaxies,
can have potentials to probe a possible evolutionary link
between nearby galaxies with inhomogeneous spatial distributions of AGB stars 
and clumpy ones at intermediate redshifts.

The purpose of this paper is to 
propose that observational studies on spatial distributions
of AGB stars for  age and metallicity ranges
in galactic disks
can probe clumpy pasts of galaxies.
By using test-particle simulations,
we investigate how unbound groups of AGB stars 
(``stellar clumps'') are dispersed
into interstellar spaces in galaxies and thereby
try to understand how long the initial inhomogeneous distributions
of stars
survive after the formation of stellar clumps. 
Since the purpose of this paper is simply to propose
the importance of 
{\it azimuthal distributions of AGB stars in the present galaxies}
in better probing  clumpy pasts of the galaxies, 
we adopt somewhat idealized models.
Our more realistic simulations on strongly bound
star clusters and stellar clumps 
in galaxies (e.g., Bekki et al. 2004a; Bekki et al. 2006) 
will provide the details on  more realistic evolution of  
star clusters and clumps in galaxies.
\begin{table}
\centering
\begin{minipage}{90mm}
\caption{Model parameters and results}
\begin{tabular}{ccccc}
(1)&(2)&(3)&(4)&(5)\\
Model no.                                  &
$r_{\rm c}$ (pc) &  
$R_{\rm c}$  (kpc)                         & 
${\sigma}_{\rm c}$ (km s$^{-1}$)               &  
$f_{\rm c}$ \\
I1 & 100 &  5.0 &  1.0  & 1.0   \\
I2 & 500  &  5.0   &  1.0 & 1.0    \\
I3 & 100  &  5.0   &  5.0 & 1.0    \\
I4 & 100 &  5.0 &  1.0  & 0.9   \\
I5 & 100 &  3.0 &  1.0  & 1.0   \\
T1 & 100 &  7.0 &  1.0  & 1.0   \\
T2 & 100 &  5.0 &  1.0  & 1.0   \\
T3 & 100 &  3.0 &  1.0  & 1.0   \\
T4 & 100 &  5.0 &  1.0  & 0.9   \\
T5 & 100 &  3.0 &  1.0  & 0.9   \\
T6 & 100 &  1.0 &  1.0  & 1.0   \\
T7 & 100 &  5.0 &  1.0  & 0.7   \\
\end{tabular}
\end{minipage}
{\bf Notes:} Cols. (1--5) model parameters: 
``I'' and ``T'' represent
isolated and tidal interaction models, respectively.
\end{table}

\section{The model}

Previous numerical simulations 
on disk galaxy formation showed that 
very massive clumps ($m_{\rm c}$)  with masses of
$10^8-10^9 {\rm M}_{\odot}$ can spiral into the nuclear
regions of galaxies owing to dynamical friction 
to finally become galactic bulges (Noguchi 1998).
Stars initially within these massive clumps 
are highly unlikely to be within the disk components of 
the present galaxies.
We thus focus on stellar clumps with $m_{\rm c} \le 10^7 {\rm M}_{\odot}$
which can be dispersed to become field stellar populations in
galaxy (and thus be observed in the present galactic disks).
Since we focus on field AGB stars in the present galaxies,
we investigate the evolution of unbound stellar clumps
that can finally become field stars after dispersal of the clumps:
The bound stellar clumps can be observed as star clusters
(i.e., not field stars) in the present galaxies
after long-term dynamical interaction with their host galaxies.
Although we show the evolution of spatial distributions
for {\it all stars} initially within clusters,  
we consider that the results can be held for 
spatial distributions for  AGB stars.

We investigate $\sim 9$ Gyr evolution of azimuthal
distributions ($N({\theta})$) of stars initially in stellar clumps
with different masses ($m_{\rm c}$), 
numbers of stars ($n_{\rm c}$), sizes ($r_{\rm c}$),
internal velocity dispersion (${\sigma}_{\rm c}$),
and orbits with respect to the centers of their host galaxies.
The initial position (${\bf x}$) and velocity (${\bf v}$) of
a clump  with respect to the center of its host galaxy 
are  ($x$,$y$,$z$) = ($R_{\rm c}$, 0,  0) 
and  ($v_{\rm x}$,$v_{\rm y}$,$v_{\rm z}$) =
(0, $f_{\rm c} V_{\rm 0}$, 0), respectively,
where $f_{\rm c}$ and $V_{\rm 0}$ are  parameters
determining the orbit and the circular velocity at $R=R_{\rm c}$,
respectively ($R$ is the distance from the galaxy center).
Each  clump is
assumed to have a  Plummer density profile
(e.g., Binney \& Tremaine 1987) and 
an isotropic velocity dispersion.

\begin{figure}
\psfig{file=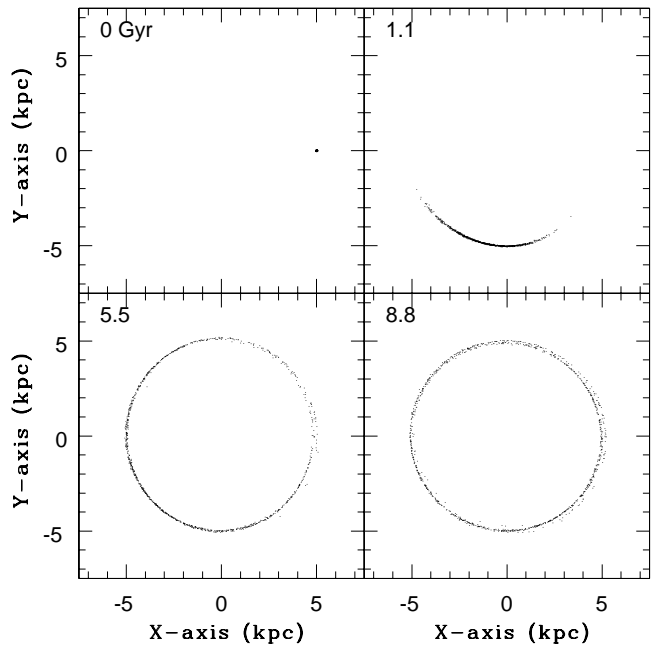,width=8.5cm}
\caption{
Time evolution of the distribution of  stars of 
an unbound star cluster (``stellar clump'') projected 
onto the $x$-$y$ plane in the standard model.
The numbers indicated in the upper left corner of each panel
represent the time ($T$ in units of Gyr)
that has elapsed since the simulation
starts. The $x$-$y$ plane corresponds to the galaxy's disk plane, which
initially includes orbital planes of stellar clumps. 
}
\label{Figure. 1}
\end{figure}

We construct models with parameters consistent with physical
properties of the LMC to compare the results with observations
for the LMC (Cioni et al. 2006a).
We adopt the same gravitational potential as used in 
previous simulations of the LMC
with the total mass  of
$2 \times 10^{10} {\rm M}_{\odot}$
(e.g., Murai \& Fujimoto 1980)
and thereby investigate the evolution of stars represented by
test-particles under the gravitational field both for
``isolated'' models  and ``tidal interaction'' ones.
The total number of AGB stars observed (e.g., Cioni \& Habing  2003)
for the LMC  
is an order of $10^4$ ($\sim 30000$), which means that the total number
of AGB stars used for each radial bin in estimating
azimuthal distributions of stars is  an order of $10^3$.
We accordingly  show the results with the models with $n_{\rm c}=10^3$.
Since we investigate test-particle simulations,  
the essential results do not depend on $n_{\rm c}$.

We mainly investigate  the isolated models in which
stellar clumps can be influenced only by
gravitational fields of their host galaxies
for $\sim 9$ Gyr.
For the tidal interaction models, we 
use the same model as used by Bekki et al. (2004b) 
and Bekki \& Chiba (2005)
in which 
orbital evolution of the LMC and the SMC
during the LMC-SMC-Galaxy interaction is investigated for
the last $\sim 9$ Gyr.
Since the typical age of AGB populations 
is several Gyrs (Cioni \&  Habing 2003), we 
mainly discuss the distributions of stars at
$T=5.5$ Gyr, where $T$ represents the time that
has elapsed since the simulation starts.

We mainly show the results of the standard   isolated  model
(referred to as  ``the standard model'' just for convenience)
with $r_{\rm c}= 100$pc, 
${\sigma}_{\rm c}=1$ km s$^{-1}$,
$R_{\rm c}=5$ kpc,  and $f_{\rm c}=1$.
Considering that (i) stellar clumps are born embedded within
giant molecular clouds  (GMCs) 
(e.g.,  Lada \& Lada 2003) and (ii) there is an observed
relationship ($m_{\rm GMC} \propto {r_{\rm GMC}}^2$)
between their masses ($m_{\rm GMC}$)
and sizes ($r_{\rm GMC}$) by Larson (1981),
we adopted the above $r_{\rm c}$  
corresponding to the sizes of GMCs  with the masses
of $10^6-10^7 {\rm M}_{\odot}$. 
The adopted ${\sigma}_{\rm c}$ corresponds 
to a typical velocity dispersion
of open clusters (e.g., Binney \& Tremaine 1987).
We shows the results of the 12 models, for which
the parameter values are given in  Table 1.
We try to quantify the degrees of inhomogeneities in the azimuthal
distributions of stars by introducing a parameter ${\sigma}_{\theta}$,
which is a dispersion in $N_{\theta}$ 
(i.e., ${{\sigma}_{\theta}}^2
=\Sigma {N_{\theta}}^2 /N_{\rm bin} - {(\Sigma N(\theta)/N_{\rm bin})}^2$,
where $N_{\rm bin}$ is the total bin number used in the present study)
for a given radius in a galaxy:
The larger (smaller) ${\sigma}_{\theta}$ 
mean higher (lower) degrees  of inhomogeneities.
We note that the  ${\sigma}_{\theta}$   statistic is only intended to
quantify the azimuthal asymmetry in these simplified models.  For real
data, a method such as tracing the mean metallicity as a function of
position (Cioni et al. 2006a) should be used.

We mainly shows the results of the models in which only a stellar clump
is orbiting a galaxy. This is  because we intend to show more 
clearly the essential ingredients
of the evolution  of  $N_{\theta}$. 
Several stellar clumps
with almost identical ages and metallicities at a given radius
can be formed in real galaxies.
Since different clumps can have different $N_{\theta}$,
it is possible that  the integrated $N_{\theta}$ of these multiple clumps 
can not clearly show an inhomogeneity.
We have investigated a model 
(``multiple clump model'') in which (i) five  clumps,
one of which has the same mass and the size of the clump
used in the standard model,  are initially located
at the same distance from a galaxy yet at different azimuthal angles,
(ii) the mass distribution follows a canonical luminosity function
of the Galactic globular cluster (Harris 1991), 
and (iii) other model parameters are the same as those in the standard model. 
We have confirmed that since the integrated  $N_{\theta}$
of the five clumps is mostly determined by the stellar distribution
of the dominant (the most massive) clump,  an inhomogeneity
in  $N_{\theta}$ can be seen in the multiple clump model.

\begin{figure}
\psfig{file=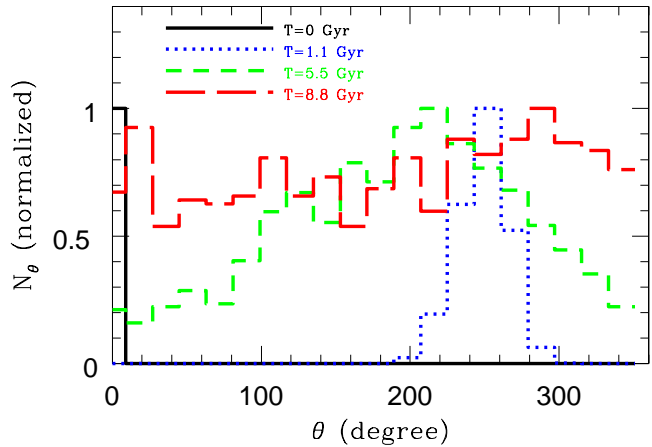,width=8.5cm}
\caption{
The azimuthal distributions ($N(\theta)$) of stars at four different
epochs, $T=0$ Gyr (black solid), $T=1.1$ Gyr (blue dotted),
$T=5.5$ Gyr (green short-dashed), and $T=8.8$ Gyr (red long-dashed),
for the standard model. For convenience, all stars at $T=0.0$
Gyr are assumed to be in the first $\theta$ bin 
(i.e., $\theta \le 10^{\circ}$). 
}
\label{Figure. 2}
\end{figure}

\begin{figure}
\psfig{file=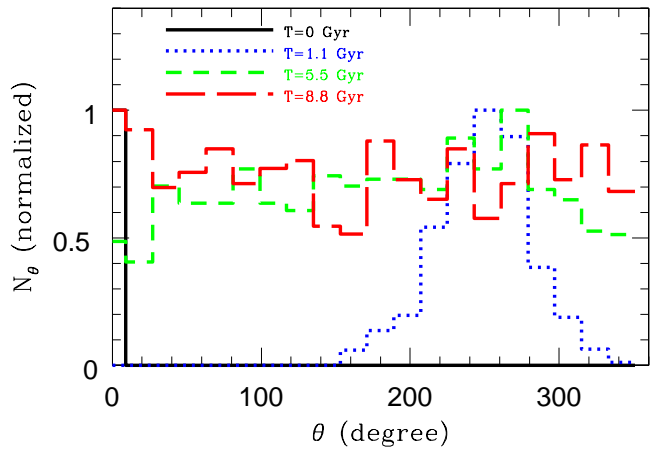,width=8.5cm}
\caption{
The same as Fig. 2 but for the model I2.
}
\label{Figure. 3}
\end{figure}

\section{Results}


Fig. 1 shows that the initial compact distribution
of the stellar clump in the standard model (I1)
is transformed into a ``crescent-shaped'' one ($T=1.1$ Gyr)
through phase mixing of stars.
As dynamical evolution proceeds, the azimuthal distribution of stars
becomes more circular and less inhomogeneous ($T=5.5$ and $8.8$ Gyr).
Fig. 2 shows that although the azimuthal distribution of
stars projected onto the $x$-$y$ plane appears to be
more  homogeneous at  $T=5.5$ Gyr (Fig. 1),
$N(\theta)$ clearly shows a high degree of inhomogeneity:
There is a factor of $\sim 5$ difference between the minimum
and the maximum $N(\theta)$.
${\sigma}_{\theta}$ is 0.27, 0.26, and 0.13 for $T$ = 1.1 Gyr,
5.5 Gyr, and 8.8 Gyr, respectively,
 which clearly indicates that the degree
of inhomogeneity in the azimuthal distribution of stars
lessens owing to phase mixing.
These results demonstrate  that the initial inhomogeneous 
distribution of stars (${\sigma}_{\theta} > 0.2$)
can be maintained in the galaxies for several Gyrs in this model.
These results thus suggest that inhomogeneous azimuthal
distributions of stars observed in the present galaxies
can probe the past clumpy stellar distributions of the 
galaxies several Gyr ago.

Fig. 3 shows that the initial inhomogeneous distribution
can much more quickly disappear in the model I2 
with a larger size of the stellar clump owing to
the more efficient phase mixing.
$N(\theta)$ is more homogeneous and thus has a smaller ${\sigma}_{\theta}$
(0.13) at $T=5.5$ Gyr in comparison with the standard model.
Considering the $m_{\rm GMC}-r_{\rm GMC}$ relation
(i.e., $m_{\rm GMC} \propto {r_{\rm GMC}}^2$)
of GMCs from where stellar clumps are assumed to form,
this result implies that initial inhomogeneities in 
the azimuthal distributions of stars can more
quickly disappear for more massive stellar clumps.
This dependence of the results on $r_{\rm c}$ does not
depend on other parameters such as ${\sigma}_{\rm c}$,$R_{\rm c}$,
and $f_{\rm c}$.

We find the  following parameter dependences
for the isolated models:
(i) the initial inhomogeneous distribution of stellar clumps
can more quickly disappear in the models with larger ${\sigma}_{\rm c}$
(${\sigma}_{\theta}$ = 0.10 at $T=5.5$ Gyr in I3),
(ii) the models with smaller $f_{\rm c}$ (i.e., more elongated orbit)
show larger ${\sigma}_{\theta}$ (e.g., 0.30 and 0.20 at $T=5.5$ Gyr
and $T=8.8$ Gyr, respectively, in I4),
and (iii) the final  ${\sigma}_{\theta}$ at $T=8.8$ Gyr does not depend
so strongly on $R_{\rm c}$ (e.g., 0.14 in I5).
These results suggest that it depends on initial properties
of stellar clumps how quickly  the initial inhomogeneous
distributions (${\sigma}_{\theta}>0.2$) disappear owing
to phase mixing in disk galaxies.
These furthermore imply that the spatial distribution
of field  stars with a given age and metallicity has a potential
to probe the initial conditions of a stellar clump
from which the  stars originate.


Fig. 4 shows that stellar clumps with different
initial parameters have different spatial distributions of stars
at $T=5.5$ Gyr because of (i) different initial conditions
of the clumps (e.g., $R_{\rm c}$) and (ii)
different strength of tidal force 
that the clumps feel from the Galaxy and the SMC. 
Fig. 5 clearly indicates that each clump has an inhomogeneous
$N(\theta)$, though the degree of inhomogeneity (${\sigma}_{\theta}$)
is different between them. 
The synthesized
azimuthal distribution for all stars from the seven models
also show a significant inhomogeneity (${\sigma}_{\theta}=0.22$).
These results imply that  inhomogeneous azimuthal
distribution of stars 
due to dispersal of stellar clumps
can last several Gyrs in strongly interacting
galaxies like the LMC and the SMC.

It is highly likely that different stellar clumps initially
have different ages and metallicities in the LMC.
The above results in Figs. 4 and 5 
accordingly imply  that a stellar clump 
with a metallicity formed about several
Gyr ago in the LMC can be probed by
an inhomogeneous azimuthal distribution of field  stars
with the same metallicity 
as obtained from the $K_{\rm s}$ magnitude distribution of AGB stars.
These results also imply that field  stars for different
metallicity ranges in the LMC 
have fossil information of different clumps formed several Gyr ago.

Thus, both isolated and tidal interaction models show that
significantly inhomogeneous azimuthal distributions of stars 
due to dispersal of unbound stellar clumps
can last several Gyrs, though
the timescales ($T_{\rm sur}$) for  the inhomogeneities
to disappear
depend strongly on physical properties of the clumps.
The self-gravity of the clumps, which is not included
in the present simulations, is highly likely
to extend the survival timescales  $T_{\rm sur}$.
We therefore suggest that the time evolution of $N(\theta)$ 
is  slower in real stellar clumps  than in
the simulated ones.

\begin{figure}
\psfig{file=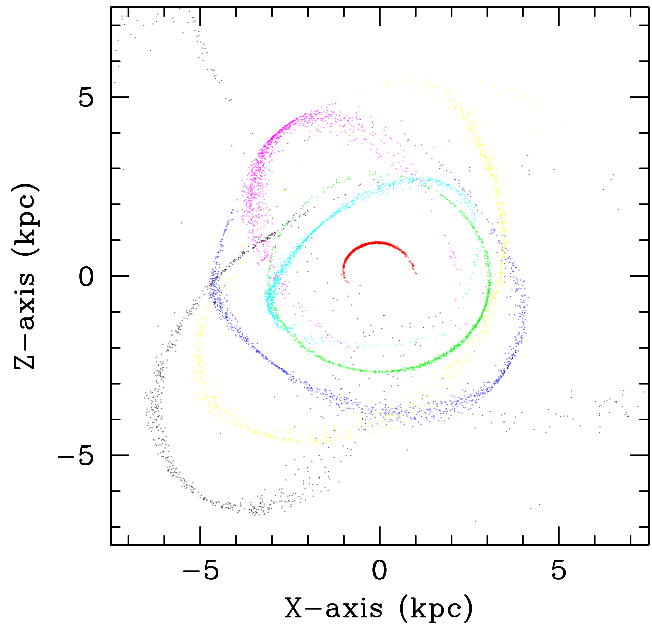,width=8.5cm}
\caption{
Distributions of stars originating from different stellar clumps
projected onto the $x$-$z$ plane for the tidal interaction models
T1$-$T7. Different colors  show different models:
black, yellow, green, blue, cyan, red, and magenta represent
T1, T2, T3, T4, T5, T6, and T7, respectively.
The initial disk defined by orbital planes of stellar clumps
in this interaction models is inclined
so that the configuration of the disk including
clumps' orbital plane (in the sky) is consistent with
observations (Bekki \& Chiba 2005).
The distribution projected  onto the $x$-$z$ plane
is similar to that viewed from the face-on of the LMC.
}
\label{Figure. 4}
\end{figure}

\begin{figure}
\psfig{file=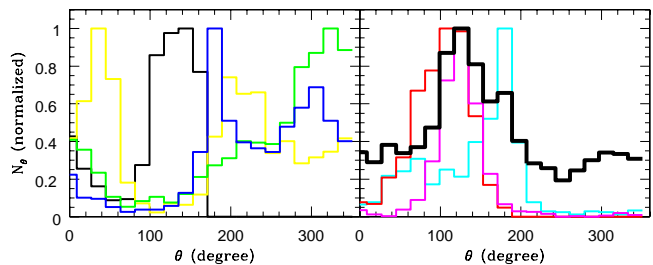,width=8.5cm}
\caption{
The azimuthal distributions ($N(\theta)$) of stars at $T=5.5$ Gyr 
for the tidal interaction models
T1$-$T7. Different colors  show different models:
black, yellow, green, blue, cyan, red, and magenta represent
T1, T2, T3, T4, T5, T6, and T7, respectively.
For comparison, $N(\theta)$ for all stars from the seven models
is shown by a thick black line.
}
\label{Figure. 5}
\end{figure}

\section{Discussions and Conclusions}

We have shown that stellar clumps formed about several Gyr ago
in the LMC's disk can leave inhomogeneous azimuthal distributions
of field  stars in the present LMC even after strong
tidal interaction between the LMC, the SMC, and the Galaxy.
This result strongly suggests that although the {\it integrated}
distribution of field  stars with different ages and metallicities
in the LMC can be relatively homogeneous,
the azimuthal  distributions of the stars {\it for a given
age and metallicity range} (thus  stars from a specific stellar clump) 
can be significantly inhomogeneous (${\sigma}_{\theta}>0.2$).
In the real LMC,  field stars
for a given radius ($R$) originate from different stellar clumps
formed with different masses and metallicities at different epochs.
The characteristic azimuthal distribution at a given radius in the LMC
accordingly would  be determined by the stars originating from
the most massive stellar clump ever formed 
around the radius.

Cioni et al. (2006a) reported that (1) mean ages
of stars are different in different regions of the LMC
and (2) mean  ages of stellar populations  can be younger  
in the northern and
southern parts of the inner disk in comparison with
the western part of the inner disk and the northern east  
part of the outer disk.
We suggest that this observed inhomogeneity can be caused
by the combination of inhomogeneous distributions of 
field AGB stars originating from massive stellar clumps
formed at different epochs: The observed inhomogeneity
suggests that the LMC had a significantly clumpy
distribution of stars about several Gyr ago.
The observed inhomogeneity in the azimuthal distribution
of metallicities of stars (Cioni et al. 2006a)
might well be closely associated with dispersal of
stars originating 
form stellar clumps with different metallicities.
It is however 
unclear whether the LMC several Gyr ago was so clumpy
as to be identified as clump-clusters (or chain galaxies).

Harris \& Zaritsky (2004) reported that the SMC has
a ring-like structure of stars with metallicities
of $Z=0.004$ and ages of about 2.5 Gyr.
The present study has demonstrated that ring-like structures
or crescent-shaped ones can be formed from stellar clumps
owing to phase mixing.
This suggests that the observed ring-like structure
in Harris \& Zaritsky (2004) is due to
the dispersal of a massive stellar clump(s)
formed about 2.5 Gyr ago: The 
observed large-scale (i.e., kpc-scale) 
annular structure does not necessarily
mean that the {\it initial} distribution of new stars in the SMC about 2.5 Gyr
was similar to a ring.

Cioni et al. (2006a) proposed that
star formation histories of AGB populations 
can be investigated by comparing 
the observed $K_{\rm s}$ magnitude distributions 
of AGB stars with the corresponding
theoretical predictions based on population
synthesis models (See also Cioni et al. 2003 for the
details of metallicity determination of AGB stars). 
Cioni et al. (2006b) investigated
the $K_{\rm s}$ magnitude distribution
of carbon stars in the SMC 
and thereby  found a ``broken-ring'' with
a larger degree of inhomogeneity in
age distributions of field stars: They confirmed
the presence of the ring reported by Harris \& Zaritsky (2004).
They also found that the stellar population
becomes increasingly more metal-rich from southern east (SE)
to northern west (NW) with increasing mean age  within the ring.
The present study suggests that 
the locations of the peak stellar densities of 
stellar clumps being dispersed (to become field stars) 
in the azimuthal direction for a given radius
are different between different clumps
(with different masses, sizes, and metallicities). 
We thus suggest that the observed inhomogeneous
age and metallicity distributions within the ring
of the SMC can be due to the dispersal of
multiple clumps with different metallicities
and ages formed at the radius where the ring
is now observed.
It is however unclear how the observed
SE-NW age and metallicity gradients were  formed 
in the SMC and what is the effect of the 
orientation of the SMC on these findings.

Massive, strongly bound stellar clumps with 
$m_{\rm c}= 10^8-10^9 {\rm M}_{\odot}$,
which could be responsible for the clumpy appearances
of galaxies such as clump-clusters at intermediate and high redshifts 
(e.g., Elmegreen et al. 2005),
are suggested to spiral into the central regions of galaxies to become
galactic bulges (e.g., Noguchi 1998).
Therefore, the presence of such  very massive clumps 
in these distant galaxies can be difficult to be  probed by 
inhomogeneous azimuthal distributions of AGB stars in {\it disk components}
of the present galaxies. 
The present study thus suggests that  
the clumpy appearances of distant  galaxies caused by
less massive clumps
can be probed by inhomogeneous azimuthal distributions of AGB stars
in the present galaxies.

AGB stars are just one of stellar populations that are formed
within star clusters, and accordingly other stellar populations
(e.g., planetary nebulae) can also probe the clumpy pasts of galaxies.
Owing to the bright magnitudes of AGB stars,
the AGB population is the most promising 
in probing the clumpy pasts not only for galaxies in the Local Group
but also for nearby ones outside the Local Group
in future observational studies.
Clumpy structures due to the presence of
bound and unbound star clusters have significant influences
on dynamical evolution of galaxies such as thick disk
formation (e.g., Kroupa 2002).
Thus probing clumpy pasts of galaxies from AGB stars
can be  important for better understanding long-term dynamical evolution
of galaxies.

\section*{Acknowledgments}
We are  grateful to the anonymous referee for valuable comments,
which contribute to improve the present paper.
K.B. acknowledges the financial support of the Australian Research
Council throughout the course of this work.


\begin{thebibliography}{}

\bibitem[]{} 
Bekki,  K., Couch,  W. J., Beasley,  M. A.,
Forbes,  D. A., Chiba, M., Da Costa G., 2004b, ApJ, 610, L93

\bibitem[]{} 
Bekki, K., Couch, W. J., Drinkwater, M. J., Shioya, Y., 
2004a, ApJ, 610, L13

\bibitem[]{} 
Bekki, K.,  Chiba, M., 2005, MNRAS, 356, 680
 
\bibitem[]{} 
Bekki, K., Couch, W. J.,  Shioya, Y., 
2006, ApJ, 642, L133

\bibitem[]{} 
Binney, J.,   
Tremaine, S., 1987 in Galactic Dynamics.

\bibitem[]{} 
Cioni, M.-R. L., Girardi, L., Marigo, P., Habing, H. J.,
2006b, A\&A, 452, 195

\bibitem[]{} 
Cioni, M.-R. L., Girardi, L., Marigo, P., Habing, H. J.,
2006a, A\&A, 448, 77

\bibitem[]{} 
Cioni, M.-R. L., Habing, H. J., 2003, A\&A, 402, 133

\bibitem[]{} 
Conselice, C. J., Grogin, N. A., Jogee, S.,
Lucas, R. A., Dahlen, T., de Mello, D.,  Gardner, J. P.,
Mobasher, B., Ravindranath, S.,
2004, ApJ, 600, L139

\bibitem[]{} 
Cowie, L. L., Hu, E. M., Songaila, A, 1995, AJ, 110, 1576

\bibitem[]{} 
Elmegreen, D. M.,  Elmegreen, B. G.,  Sheets, C.  M., 2004a, ApJ, 603, L74

\bibitem[]{} 
Elmegreen, D. M.,  Elmegreen, B. G.,  Hirst, A. C., 2004b, ApJ, 604, L74

\bibitem[]{} 
Elmegreen, D. M.,  Elmegreen, B. G., Rubin, D. S.,
Schaffer, M. A., 2005, ApJ, 631, 85


\bibitem[]{} 
Harris, J.,  Zaritsky, D., 2004, AJ, 127, 1531

\bibitem[]{} 
Harris, W. E., 1991, ARA\&A, 29, 543

\bibitem[]{} 
Kroupa, P., 2002, MNRAS, 330, 707

\bibitem[]{} 
Lada, C. J., Lada, E. A., 2003, ARA\&A, 41, 57

\bibitem[]{} 
Larson, R. B., 1981, MNRAS, 194, 809

\bibitem[]{} 
Murai, T., Fujimoto, M., 1980, PASJ, 32, 581

\bibitem[]{} 
Noguchi, M., 1998, Nat, 392, 253


\bibitem[]{} 
Reshetnikov, V. P., Dettmar, R.-J.,  Combes, F., 2003, A\&A, 399, 879

\bibitem[]{} 
van den Bergh, S., Abraham, R. G., Ellis, R. S.,
Tanvir, N. R., Santiago, B. X.,  Glazebrook, K. G.,
1996, ApJ, 463, 602

\bibitem[]{} 
van der Marel, R.  P., Alves, D. R., Hardy, E.,  Suntzeff, N. B.,
2002, AJ, 124, 2639

\end{thebibliography}
\end{document}